\documentclass{aa}  
\usepackage{graphicx}
\usepackage{txfonts}
\usepackage{natbib}
\bibpunct{(}{)}{;}{a}{}{,}
\begin{document}

  \title{Synthetic NLTE accretion disc spectra for the dwarf nova SS Cyg during an outburst cycle}
  \titlerunning{NLTE accretion disc spectra for SS Cyg}
  \author{M. Kromer\inst{1,2}, T. Nagel\inst{1}, \and K. Werner\inst{1}}
  \offprints{T. Nagel,\\
             \email{nagel@astro.uni-tuebingen.de}
            }
  \institute{Institut f\"ur Astronomie und Astrophysik,
             Universit\"at T\"ubingen,
             Sand 1, 72076 T\"ubingen, Germany
\and
Max-Planck-Institut f\"ur Astrophysik, Karl-Schwarzschild-Stra\ss e 1, 85741 Garching, Germany
            }

  \date{Received xxxx; accepted xxxx}

  \abstract
  {Dwarf nova outbursts result from enhanced mass transport through the accretion
disc of a cataclysmic variable system.}
  {We assess the question of whether these outbursts are caused by an enhanced
mass transfer from the late-type main sequence star onto the white dwarf
(so-called mass transfer instability model, MTI) or by a thermal instability in
the accretion disc (disc instability model, DIM).}
  {We compute non-LTE models and spectra of accretion discs in quiescence and
outburst and construct spectral time sequences for discs over a complete
outburst cycle. We then compare our spectra to published optical spectroscopy of the
dwarf nova SS~Cygni. In particular, we investigate the hydrogen and helium line
profiles that are turning from emission into absorption during the rise to
outburst.}
  {The evolution of the hydrogen and helium line profiles during the rise to outburst
and decline clearly favour the disc-instability model. Our spectral model
sequences allow us to distinguish inside-out and outside-in moving heating waves in
the disc of SS~Cygni, which can be related to symmetric and asymmetric outburst
light curves, respectively.}
  {}

  \maketitle
%

\section{Introduction}

  Dwarf novae (DN) belong to the non-magnetic cataclysmic variables which
  are binary systems consisting of a white dwarf as primary component and
  an orbiting late-type main sequence star. Due to their close orbit, mass
  transfer from the secondary onto the primary via Roche lobe overflow
  occurs. Because of conservation of angular momentum an accretion disc
  forms around the white dwarf (Warner 1995). 

  Dwarf novae are characterised by more or less regular outbursts during
  which the system undergoes a rise in optical brightness of 2-6 magnitudes. The
  observed outbursts can be divided into two categories, depending whether
  the lightcurves are symmetric or not. For asymmetric outbursts they are
  characterised by a fast rise and slower decline and show a delay of the
  rise in the UV against the optical. For symmetric outbursts UV and
  optical fluxes rise simultaneously on a longer timescale than the
  asymmetric outbursts. Sometimes for both types plateaus are observed
  during maximum. It is commonly accepted that the outbursts are caused by
  a luminosity increase in the disc that arises from a temporarily
  increased mass transport through the disc. The origin of this increased
  mass transport, however, has been discussed controversially. 

  According to the mass transfer instability model (MTI, Bath 1975)
  an instability in the secondary star leads to a temporarily increased
  mass transfer rate $\dot{M}_2$ from the secondary onto the disc, so that
  also the mass transport through the disc increases. Because the
  instability originates in the secondary, the outbursts must start at the
  outer edge of the disc and proceed inwards in the framework of this
  model. 

  In the disc instability model (DIM, Osaki 1974) the mass transfer
  from the secondary is constant and the outbursts are attributed to
  thermal viscous instabilities in the disc that lead to a temporarily
  increased mass transport through the disc. Meyer \& Meyer-Hofmeister
  \cite{Meyer1981} and Faulkner et al. \cite{Faulkner1983} have shown that
  this instability is due to the local ionisation of hydrogen. Radial
  temperature and viscosity gradients lead to the propagation of heating or
  cooling waves throughout the disc, which carry the whole disc over to
  outburst or quiescence, respectively. In particular this allows the
  outbursts to start any place in the disc so that outbursts can proceed
  inwards or outwards.  

  Today the DIM is generally favoured over the MTI. This is due to the
  existence of a detailed theoretical framework for the DIM that can
  explain the outburst behaviour and the different outburst types in a
  natural manner. According to Smak \cite{Smak1984a} the different rise times
  for asymmetric and symmetric outbursts are caused by different
  propagation directions of the outbursts: for the asymmetric outbursts
  (type-A after Smak 1984a) the heating wave originates in the
  outer part of the disc and proceeds inside moving with the mass stream,
  so that the heating wave can move relatively fast. Finally the hot inner
  part of the disc switches to outburst, so that the rise in the UV is
  delayed against the optical. In contrast to the symmetric type-B
  outbursts, the heating wave proceeds inside-out and therefore has to move
  against the mass stream, resulting in a relatively slow rise. In this case
  the hot inner parts of the disc are the first to switch to outburst so
  that optical and UV fluxes increase simultaneously. 

  Furthermore, that no luminosity increase in the hotspot - the
  region of the disc, where the mass stream from the secondary impinges -
  is observed during an outburst contradicts the MTI model, because such a
  luminosity increase would be expected if the mass transfer from the
  secondary increases.  

  A possibility of distinguishing between both models by observational data
  arises if one is able to decide whether the outbursts proceed outside-in
  or inside-out. This can be achieved by comparing time-resolved spectra
  for an entire outburst cycle with the appropriate model spectra because
  quiescence and outburst spectra differ significantly. In quiescence the
  optical spectrum shows the strong hydrogen Balmer emission lines characteristic
  of an optically thin disc. In contrast, during outburst, broad absorption
  features in the Balmer series indicate an optically thick disc. At the
  same time, the intensity in the blue wavelength range increases
  particularly strongly indicating a rise in the disc temperature. 

  To this end we calculated time-resolved model spectra (Sect.~3) tailored
  to the dwarf nova SS Cyg. This is the brightest known DN, showing an
  outburst brightness of \(V=8.2\,\mathrm{mag}\) (Ritter \& Kolb 2003),
  making it one of the best-studied DN. Before presenting these models
  in Sect.~3, we give a short overview of our approach in Sect.~2. The
  results are discussed in Sect.~4.   

\section{Model assumptions}

  To calculate the accretion disc spectra we use our accretion disc code
  AcDc (Nagel 2003; Nagel et al. 2004), which is based on the assumption of
  a geometrically thin disc (total disc thickness \(H\) is much smaller
  than the disc diameter). This allows us to decouple vertical and radial
  structures and, together with the assumption of axial symmetry, to
  separate the disc into concentric annuli of plane-parallel geometry. Then
  radiative transfer becomes a one-dimensional problem.   

  Each of these disc rings, which are located at a given radial distance
  \(r\) from the white dwarf, is assumed to be stationary. Thus it can be
  characterised by a constant mass transport rate \(\dot{M}\). The rate of
  energy generation from viscous shear then becomes independent of the
  kinematic viscosity \(\nu_\mathrm{k}\) and can be parameterised by the
  effective temperature 
  \label{eq:Teff}
  \begin{equation}
    T_\mathrm{eff}(r) = \left[ \frac{3GM_1\dot{M}} {8\pi\sigma r^3} 
                            \left(1 - \sqrt{\frac{r_1} {r}}\right)
                       \right]^{1/4}
  \end{equation}
  (for example Warner 1995). Thereby $M_1$ denotes the mass, \(r_1\) the
  radius of the primary white dwarf, \(G\) the gravitational constant, and
  \(\sigma\) the Stefan-Boltzmann constant.  

  To get a self-consistent solution, the radiative transfer equation, the hydrostatic and
  energy equilibrium equations, as well as the NLTE rate equations, that determine
  the occupation numbers of the atomic levels, are solved simultaneously by
  an iterative scheme. Therefore detailed information about the involved
  atomic levels is needed, which is provided in the form of a model atom
  (cf. Rauch \& Deetjen 2003). The kinematic viscosity, which is needed for
  the vertical structure calculation, can be parameterised by the \(\alpha\)-approach
  of Shakura \& Sunyaev \cite{Shakura1973}
  \begin{equation}
    \nu_\mathrm{k}=\alpha c_\mathrm{s} H
  \end{equation}
  (where \(c_\mathrm{s}\) is the speed of sound and \(\alpha\leq1\) a
  dimensionless parameter) or after Lynden-Bell \& Pringle
  \cite{Lynden-Bell1974} by the Reynolds number $Re$ 
  \begin{equation}
    \nu_\mathrm{k}=\frac{r v_\phi}{Re} = \frac{\sqrt{G M_1 r}}{Re}\,,
  \end{equation}
  where \(v_\phi\) is the Kepler velocity. We choose the latter approach
  that is numerically easier to implement because we save a further iteration to
  solve consistently for \(c_\mathrm{s}\) and \(H\). 

  Irradiation of the disc by the primary is considered via the upper
  boundary condition for the radiative transfer equation. For that purpose
  the irradiation angle \(\beta\) for each disc ring and the spectrum of
  the primary must be specified. The spectrum of the primary is
  parameterised by a blackbody temperature \(T_\mathrm{bb}\), or detailed
  white dwarf model atmosphere spectra are calculated. 

  The complete set of input parameters, which we must provide for each disc
  ring, thus consists of \(M_1, r_1, \dot{M}, r, Re, \beta,
  T_\mathrm{bb}\). The spectrum of the complete disc is then obtained by
  integrating the spectra of these disc rings for different inclination
  angles, the spectral lines are Doppler shifted according to the
  radial component of the Keplerian rotation velocity in the disc. 

  As the accretion discs of dwarf novae are fed by a late-type main
  sequence star, we assume a disc composition of hydrogen and helium with
  relative solar abundances. The model atoms used for the disc model
  calculations presented here contain the ionisation stages \ion{H}{i},
  \ion{H}{ii}, and \ion{He}{i}-\ion{He}{iii}. The number of NLTE levels and
  lines considered is 15 and 105 for \ion{H}{i}, 29 and 61 for \ion{He}{i},
  and 14 and 78 for \ion{He}{ii}. We consider the \(\mathrm{H}^{-}\)
  opacity and Rayleigh scattering for H and He, which is important for
  the coolest regions of the disc model. 
In addition, the Ly$_\alpha$ line in the cool models for the quiescent
  disk is so broad that it contributes considerably to the source function
  in the optical band. The reason is that most of the hydrogen (about
  90\,-\,99\%) is neutral throughout most of the line-forming region
  (Fig.~\ref{fig:13.5hot_vstruct}, second panel from top).

  \section{Models}
  In the following we present detailed models for the accretion disc of SS
  Cyg in outburst and quiescence. 
  SS Cyg is the brightest known DN and belongs to the U Gem type of DN. 
  For our models we have chosen the orbital parameters according to
  Ritter \& Kolb \cite{Ritter2003}, who give \(M_1 =
  (1.19\pm0.02)$~$\mathrm{M}_\odot\) for the mass of the white
  dwarf. According to the mass-radius relation, this corresponds to a white
  dwarf radius of \(3.9\cdot10^8\,\mathrm{cm}\). Together with the mass of
  the companion \(M_2 = (0.704\pm0.002) $~$\mathrm{M}_\odot\) and the
  orbital period \(P = 6.6031\,\mathrm{h}\), the tidal radius -- the
  radius where the disc is disrupted by tidal interactions with the
  secondary -- follows from  
  \begin{equation} 
    r_\mathrm{tidal} =0.60\cdot\frac{a}{1+q} 
  \end{equation}
  (Hellier 2001) and amounts to \(r_\mathrm{tidal} =
  5.78\cdot10^{10}\, \mathrm{cm}\). Here \(a\) denotes the distance between
  primary and secondary, and it can be calculated from the third Kepler
  law. \(q\) is the mass ratio \(\frac{M_2}{M_1}\). The minimal extension
  of the disc is given by the so-called circularisation radius  
  \begin{equation}
    r_\mathrm{circ} = r^4_{\mathrm{L}_1}\cdot\frac{1+q}{a^3} 
  \end{equation}
  at which the angular momentum is equal to the angular momentum at the
  Lagrange point \(\mathrm{L}_1\). The Roche lobe and therefore the
  distance \(r_{\mathrm{L}_1}\) of the Lagrange point \(\mathrm{L}_1\) must
  be calculated numerically. After Plavec \& Kratochvil \cite{Plavec1964},
  however, for \(0.1<q<10\) the approximation 
  \begin{equation}
    r_{\mathrm{L}_1} = a\cdot\left( 0.500-0.227\log q\right)
  \end{equation}
  is possible. This finally leads to \(r_\mathrm{circ} = 1.65\cdot10^{10}\,
  \mathrm{cm}\). In this radial range we have increased the disc's outer
  edge \(r_\mathrm{o}\) until the double-peaked line profiles matched the
  observation, so we chose \(r_\mathrm{o} = 4\cdot10^{10}\,\mathrm{cm}\). A
  lower value would give line profiles that are too broad due to the higher Kepler
  rotation velocity for smaller radii.   

  The inner edge of the disc model was fixed from the following
  arguments. Our model cannot be applied to the boundary layer expected at
  the transition from the disc to the primary, so we must truncate the disc
  well before the white dwarf. Despite the higher temperatures in the inner
  disc, this has little influence on the optical spectrum because the
  surface area of the inner parts of the disc is much smaller than that of
  the outer parts. Thus the inner parts can be neglected for modelling
  optical spectra. In contrast, in the UV range there will be strong
  imprints from the inner disc portions (cf. Fig.~\ref{fig:hotrings}). 

    \begin{table}[b]
      \caption{Parameters of the rings for the hot disc in SS~Cyg.} 
      \label{tab:SSCyg_hotdiscparas}
      \begin{center}
      \begin{tabular}{rrrrrr}
        \hline\hline\noalign{\smallskip}
        \# & $r\left[10^{9}\,\textrm{cm}\right]$ & $Re$ & $T_{\mathrm{eff}}\left[\textrm{K}\right]$ & $\tau_{\mathrm{tot}}$ &  $h\left[10^{8}\,\textrm{cm}\right]$\\
        \noalign{\smallskip}\hline\noalign{\smallskip}
        1  &  1.00 & 3200 &  74912 & 104 & 0.21\\
        2  &  1.10 & 3200 &  71056 & 110 & 0.23\\
        3  &  1.22 & 3000 &  66935 & 114 & 0.26\\
        4  &  1.35 & 2900 &  63013 & 119 & 0.30\\
        5  &  1.50 & 2800 &  59074 & 124 & 0.34\\
        6  &  1.66 & 2600 &  56742 & 125 & 0.39\\
        7  &  1.84 & 2500 &  51916 & 127 & 0.44\\
        8  &  2.05 & 2400 &  48403 & 128 & 0.50\\
        9  &  2.30 & 2350 &  44873 & 130 & 0.58\\
        10 &  2.60 & 2300 &  41350 & 131 & 0.67\\
        11 &  2.97 & 2300 &  37798 & 134 & 0.79\\
        12 &  3.43 & 2300 &  34259 & 139 & 0.93\\
        13 &  4.02 & 2200 &  30705 & 145 & 1.13\\
        14 &  4.80 & 2100 &  27135 & 155 & 1.40\\
        15 &  5.85 & 1700 &  23610 & 163 & 1.81\\
        16 &  7.35 & 1620 &  20080 & 192 & 2.38\\
        17 &  9.65 & 1550 &  16525 & 221 & 3.29\\
        18 & 13.50 & 1450 &  12969 & 258 & 4.93\\
        19 & 21.00 & 1200 &   9404 & 169 & 8.14\\
        20 & 40.00 &  500 &   5862 & 1.61 & 16.18\\
        \noalign{\smallskip}\hline
      \end{tabular}
      \end{center}
      \begin{quote}
        $\tau_{\mathrm{tot}}$ is the total Rosseland optical depth from top
        to disc midplane and $h=H/2$ the vertical extension of the disc from the
        midplane. The other symbols are defined in the text.
      \end{quote}
    \end{table}

  For the inclination angle $i$ we have chosen \(40\degr\), which is
  consistent with the value of \(i=\left(37\pm5\right)\degr\) given by
  Ritter \& Kolb \cite{Ritter2003}. All disc rings have been irradiated with a
  \(50\,000\,\mathrm{K}\) blackbody spectrum. This temperature is
  compatible with the observational results for the WD in SS~Cyg
  (Long et al. 2005; Smak 1984b). Tests with a
  \(50\,000\,\mathrm{K}\) white dwarf model atmosphere have shown that the
  blackbody approximation has little influence on the emerging disc spectra. The
  irradiation angle was set to \(1\degr\). According to the system geometry, this is possible but
  probably marks an upper limit. For such small angles, the irradiation
  increases the effective temperature of the disc ring compared to the
  value expected according to Eq.~\ref{eq:Teff} only
  marginally. The relative differences are below \(10^{-3}\) and decrease with
  increasing \(r\). 

    \subsection{Outburst}

    For the hot disc during outburst, we assume a constant mass transport rate
    through the disc of \(\dot{M}=4\cdot10^{-9}~\mathrm{M}_\odot/\mathrm{yr}\)
    and a constant viscosity of \(\alpha\approx0.30\) according to
    the DIM. With those parameters, we calculated a disc model from
    \(1\cdot10^9\,\mathrm{cm} \leq r \leq 4\cdot10^{10}\,\mathrm{cm}\) by
    dividing the disc into 20 rings to obtain a smooth  distribution of
    \(T_\mathrm{eff}(r)\) with a maximal difference of \(\sim
    3500\,\mathrm{K}\) between neighbouring rings
    (see Table~\ref{tab:SSCyg_hotdiscparas}). The resulting disc is optically
    thick except for the outermost  ring. This ring's effective temperature
    of \(5862\,\mathrm{K}\) is more typical for a cold disc.

    \begin{figure}
      \centering
\includegraphics[width=0.90\columnwidth]{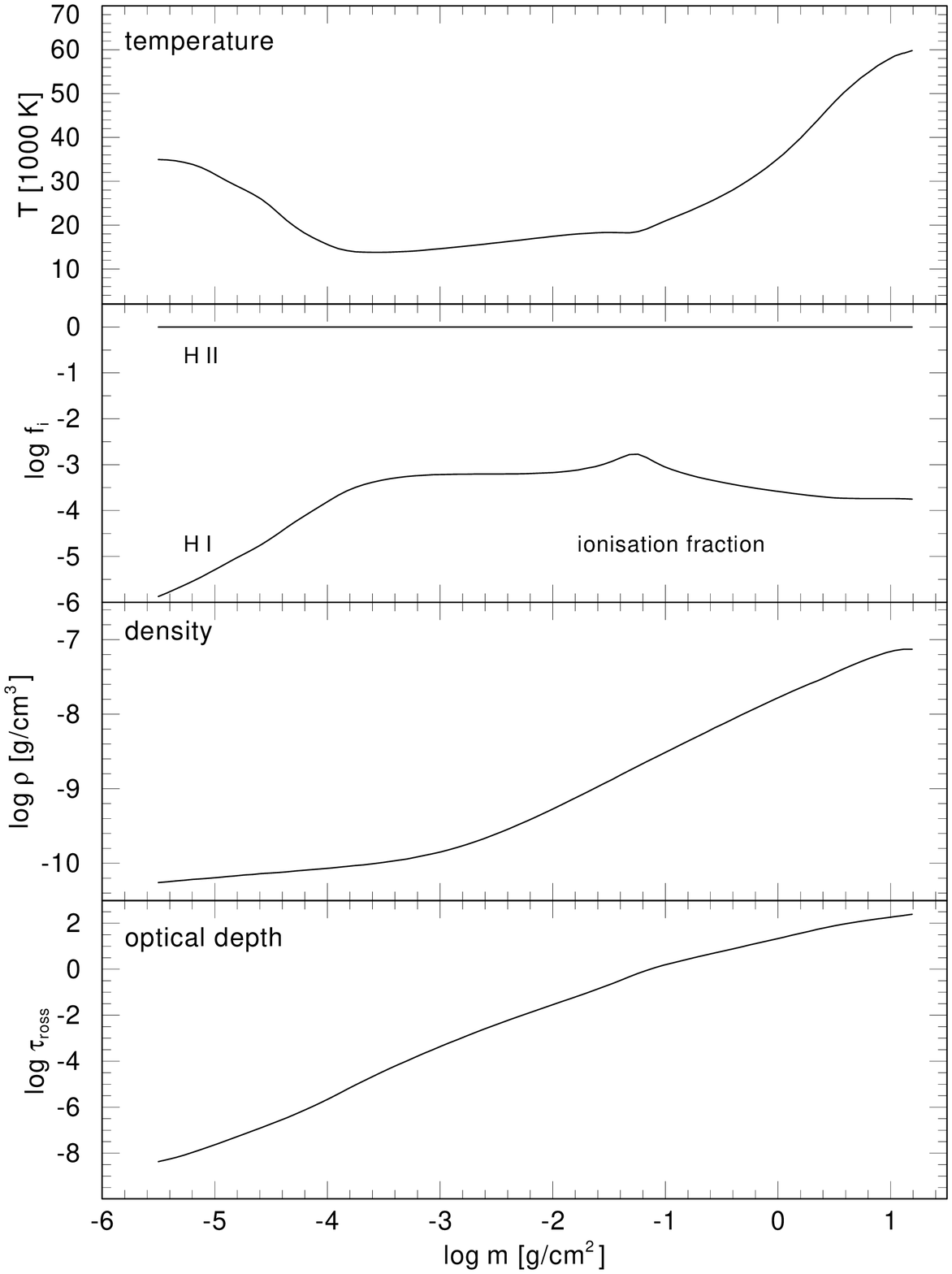}
      \caption{Vertical structure of the hot disc at a distance of
      \(7.35\cdot10^9\,\mathrm{cm}\) from the white dwarf. 
      The physical variables are plotted against the column mass measured
      from the surface towards the midplane.} 
      \label{fig:13.5hot_vstruct}
    \end{figure}

    \begin{figure*}
      \centering
\includegraphics[width=0.80\textwidth]{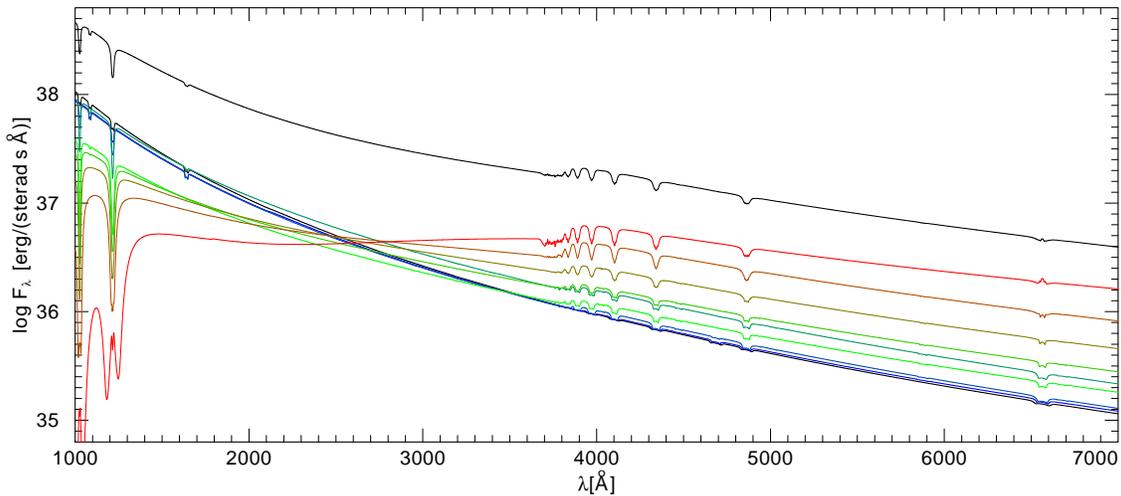}
      \caption{Model spectra for the accretion disc of SS Cyg in outburst
        (uppermost curve). The other curves show the contribution of
        selected individual disc rings starting with the outermost ring (top) and then
        continuing towards the inner disc edge. The inclination angle is
        \(40\degr\).} 
      \label{fig:hotrings}
    \end{figure*}

    As an example of the vertical structure of the disc model, the
    temperature, hydrogen ionisation fraction, density, and optical depth are plotted against the
    height above the disc midplane in Fig.~\ref{fig:13.5hot_vstruct} 
    at a distance of \(7.35\cdot10^9\,\mathrm{cm}\) from the white
    dwarf. The temperature shows an inversion at the disc surface due to
    the heating by irradiation of the WD before a strong drop down at
    \(\log m\approx-4\) occurs and then the temperature rises slowly
    towards the disc's midplane. At \(\log m \approx -1\) the disc becomes
    optically thick. 

    Figure~\ref{fig:hotrings} shows the integrated spectrum of the hot disc
    and the contribution of selected disc rings. Doppler broadening due to the
    Keplerian velocity is taken into account. In the optical it is
    characterised by hydrogen Balmer absorption lines.
    In the UV, strong absorption lines of the hydrogen Lyman series and
    \ion{He}{ii} appear. The latter lines originate in the inner disc rings,
    where \(T_\mathrm{eff}\) becomes high enough to populate \ion{He}{ii}
    levels. This high \(T_\mathrm{eff}\) is also the reason the
    inner disc rings dominate the spectrum in the UV range despite their
    small surface area compared to the outer rings. The situation is
    completely different in the optical. There the disc spectrum is dominated 
    by the outermost ring due to its large surface area, so the weak
    absorption line of \ion{He}{ii} at \(4686\,\AA\), which is visible in
    the inner ring spectra, is outshone by the much larger continuum flux
    of the outer rings.  

\begin{figure}
      \centering
      \resizebox{\hsize}{!}{\includegraphics{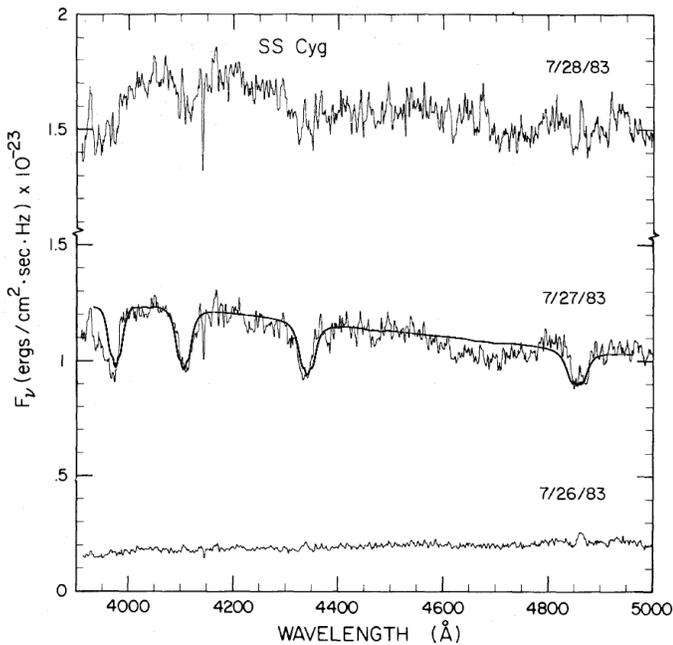}}
      \caption{Observed spectra of SS Cyg (from Clarke et al. 1984) during
      rise to outburst (outburst maximum corresponds to
      7/28/83). Overplotted is our synthetic spectrum for the accretion
      disc in outburst. The model flux was
      multiplied by a constant factor to match the observed continuum flux.} 
      \label{fig:SSCyg_outburst}
    \end{figure}

    For a comparison to observational results, Fig.~\ref{fig:SSCyg_outburst}
    shows our synthetic spectrum of the hot accretion disc with the spectra of
    Clarke et al. \cite{Clarke1984}, who observed SS~Cyg during a rise to outburst.  

    \subsection{Quiescence}
    Wood et al. \cite{Wood1986} studied the radial temperature distribution in the
    accretion disc of the DN Z Cha by eclipse mapping. In contrast to the
    \(T_\mathrm{eff} \propto r^{-3/4}\) power law expected for stationary
    accretion discs, they found a more or less constant value of the
    effective temperature at a level of several thousand Kelvin. This has
    been interpreted as a hint that the accretion discs of DN in quiescence
    are not stationary. Therefore we assumed a constant effective
    temperature of $\sim 4200\,\mathrm{K}$ throughout the 
    disc, which is in the typical range for cold discs. To achieve this temperature for all
    rings, we adjusted the mass transport rates through the rings (see
    Table~\ref{tab:SSCyg_colddiscparas}). We also adjusted the Reynolds
    number to get typical values for $\alpha$ in a disc in quiescence according to
    the DIM. The resulting kinematic viscosity is smaller than in the hot
    disc by a factor of 10, except for the outermost ring, where the
    viscosity is as high as for disc rings in outburst. For this disc ring,
    it was not possible to construct a low-viscosity model with strong emission
    lines.

    \begin{figure*}
      \centering
\includegraphics[width=0.80\textwidth]{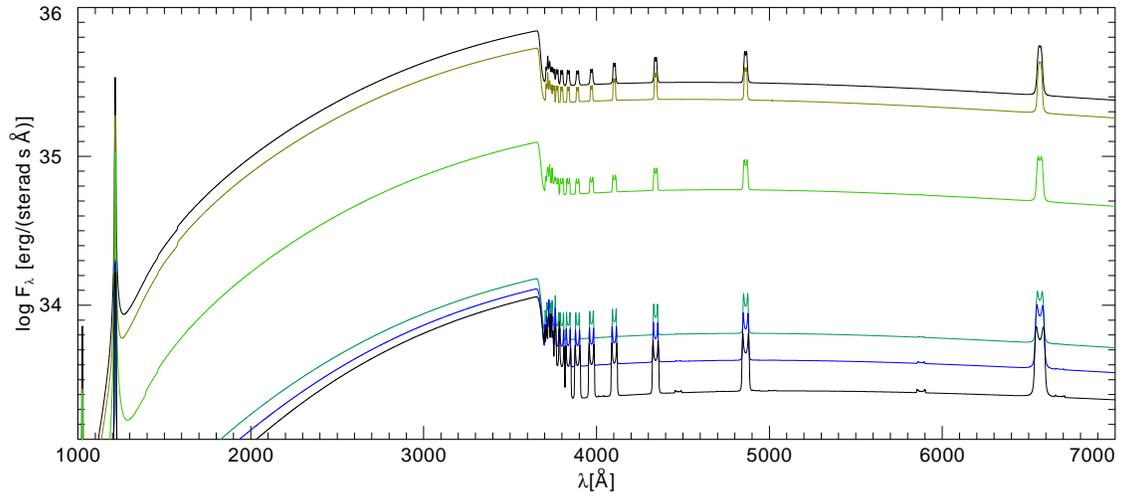}
      \caption{Model spectra for the accretion disc of SS~Cyg in quiescence
        (uppermost curve). The other curves show the contribution of selected
        individual disc rings starting with the outermost ring (top) and then
        continuing towards the inner disc edge. The inclination is
        \(40\degr\). The spectral lines are getting broader due to the
        higher Kepler velocities of the inner disc rings.} 
      \label{fig:coldrings}
    \end{figure*}

    \begin{table}[h]
      \caption{Parameters for the ring models of the cold disc in SS Cyg.} 
      \label{tab:SSCyg_colddiscparas}
      \centering
      \begin{tabular}{rrlrrr}
        \hline\hline\noalign{\smallskip}
        \# & $r\left[10^{9}\,\mathrm{cm}\right]$ & $\dot{M}\left[{\mathrm M}_\odot/\mathrm{yr}\right]$ & $Re$ & $\tau_{\mathrm{tot}}$ & $h\left[10^{8}\,\mathrm{cm}\right]$\\
        \noalign{\smallskip}\hline\noalign{\smallskip}
        1  & 4.00 & \(1.4\cdot10^{-12}\)  & 19000 &  0.27 &  0.71\\
        2  & 6.00 & \(4.0\cdot10^{-12}\)  & 16000 &  0.24 &  1.13\\
        3  & 8.00 & \(1.0\cdot10^{-11}\)  & 13000 &  0.28 &  1.59\\
        4  & 9.00 & \(1.4\cdot10^{-11}\)  & 13000 &  0.30 &  2.00\\
        5  & 10.00 & \(1.9\cdot10^{-11}\) & 10000 &  0.29 &  2.34\\
        6  & 20.00 & \(1.4\cdot10^{-10}\) &  3000 &  0.28 &  6.00\\
        7  & 40.00 & \(1.0\cdot10^{-9}\)  &   500 &  0.22 & 14.56\\
        \noalign{\smallskip}\hline
      \end{tabular}
    \end{table}

The number of disc rings required is much
    smaller than for the hot disc, as the change in spectral properties
    across the radius is marginal due to the constant temperature.  
    Furthermore we did not extend the cold disc as far in as the hot disc,
    but truncated the model at \(r=4\cdot10^9\,\mathrm{cm}\). This is again
    justified by the constant effective temperature, which prevents a strong
    contribution of the inner rings to the UV-flux in contrast to the case
    of the hot disc (see Fig.~\ref{fig:coldrings}). 
    \begin{figure}
      \centering
      \resizebox{\hsize}{!}{\includegraphics{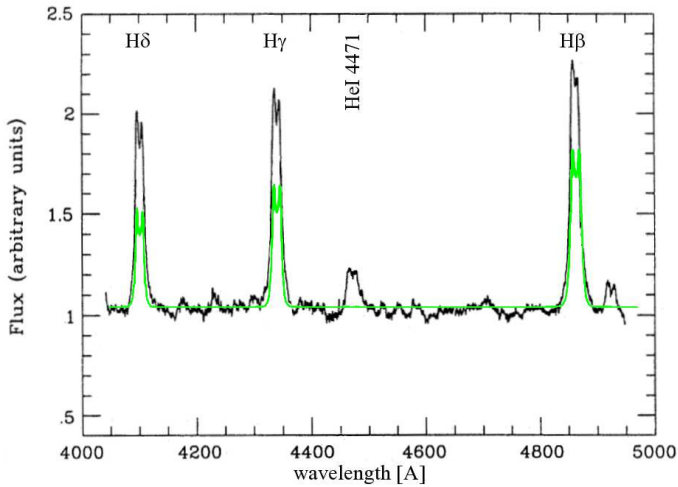}}
      \caption{Observed spectra of the accretion disc of SS Cyg during
      quiescence (Martinez-Pais et al. 1994). Overplotted is our synthetic
      spectrum for the accretion disc of SS Cyg in quiescence (grey). The flux was
      normalised to the local continuum flux.} 
      \label{fig:SSCyg_quiescence}
    \end{figure}

    The resulting disc spectrum (Fig.~\ref{fig:coldrings}) is compared to
    an observed spectrum taken from Martinez-Pais et al. \cite{Martinez-Pais1994} in
    Fig.~\ref{fig:SSCyg_quiescence}. For that purpose the model spectrum
    has been normalised to the local continuum flux. In principle, the model
    reproduces the hydrogen Balmer emission lines, but they are not as
    strong as in the observation. The \ion{He}{i} emission lines are not
    seen in our model spectrum, and the reason may be that they form in the hot
    spot, which we have not included in our models.
 
\begin{figure}
      \centering
\includegraphics[width=0.90\columnwidth]{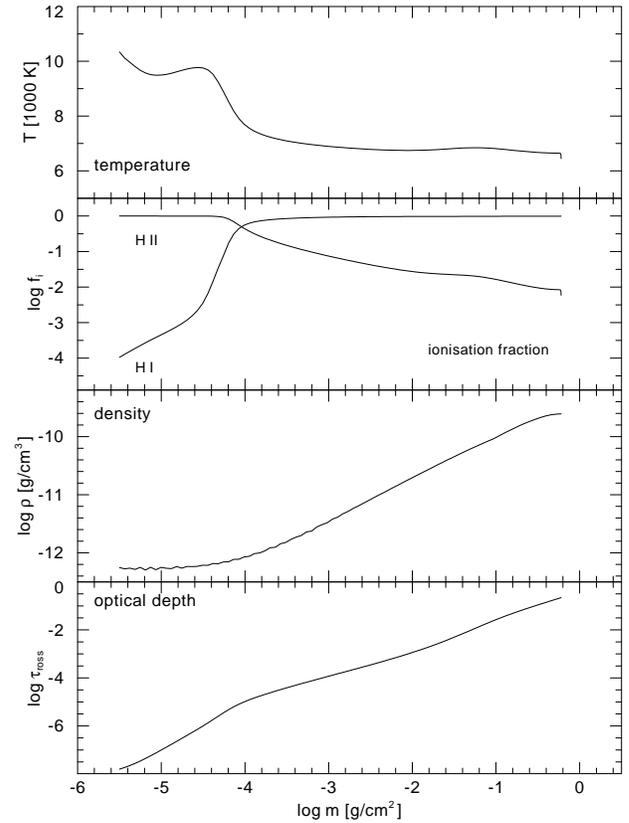}
      \caption{Vertical structure of the cold disc at a distance of
      \(4.0\cdot10^{10}\,\mathrm{cm}\) from the white dwarf. 
      The physical variables are plotted against the column mass measured
      from the surface towards the midplane. }  
      \label{fig:13.5cold_vstruct}
    \end{figure}

    As an example, the vertical structure of the cold disc is shown in
    Fig.~\ref{fig:13.5cold_vstruct} at a radial distance of
    \(4.0\cdot10^{10}\,\mathrm{cm}\). In contrast to the hot disc, the
    temperature does not increase towards the disc midplane but declines
    monotonically. Towards the disc surface, the irradiation of the WD again causes
    a temperature inversion. The entire cold disc is optically thin.

    \subsection{Rise to outburst}
    To examine the spectral evolution from quiescence to outburst, we combined
    rings of the cold and hot discs to a sequence of disc models such that this
    sequence simulates the propagation of the heating wave throughout the
    disc. In this way we studied the two different cases of outside-in and
    inside-out moving heating waves to achieve further insight into the
    processes taking place in the disc during rise to outburst. 
    For the outside-in outburst this sequence consists of five disc models
    in which the cold rings have been replaced by the next-neighbouring hot
    rings from the outside. The assembly of these models is shown in
    Table~\ref{tab:outside-in}. 

\begin{figure*}
      \centering
\includegraphics[width=12cm]{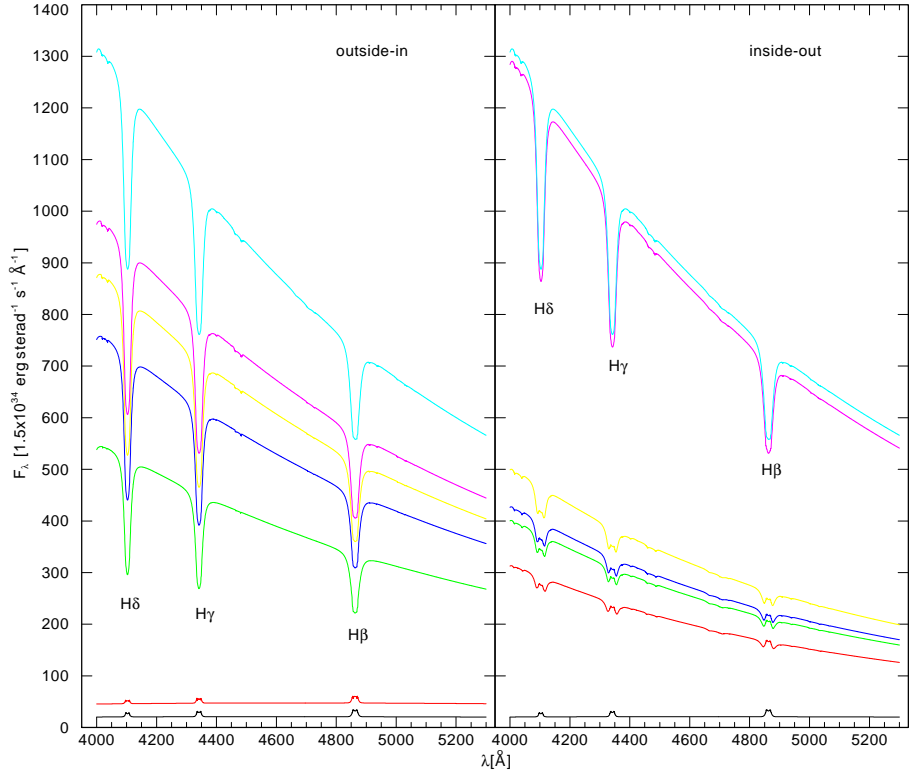}
      \caption{Spectral evolution between \(4000\) and \(5300\,\AA\) for an
        outside-in outburst (left panel) and an inside-out outburst (right
        panel). The lowermost graphs show a pure cold disc that evolves to
        full outburst (uppermost graph) over steps 1 to 5 of
        Tables~\ref{tab:outside-in} and \ref{tab:inside-out},
        respectively.} 
      \label{fig:outburstsequence}
    \end{figure*}

\begin{figure*}
      \centering
\includegraphics[width=12cm]{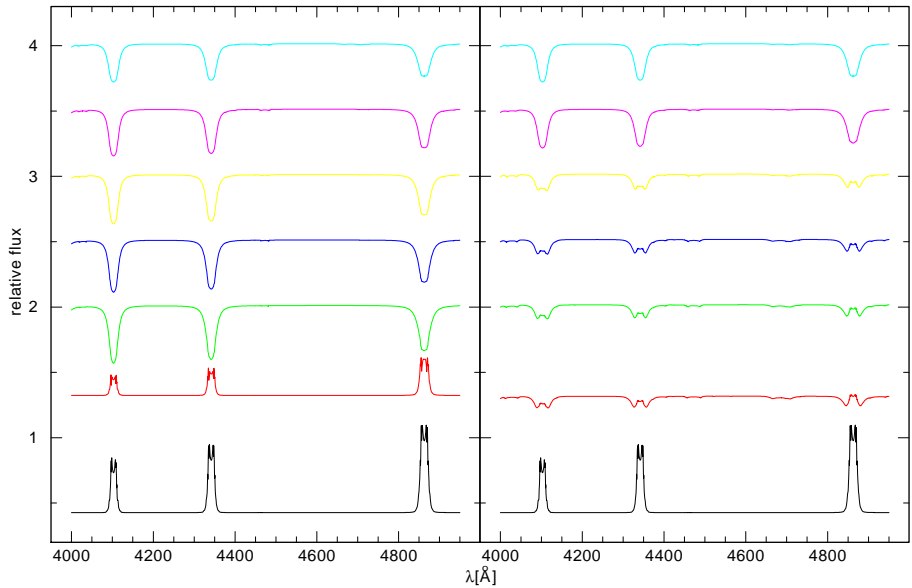}
      \caption{Spectral evolution between \(4000\) and \(5300\,\AA\) for an
        outside-in outburst (left panel) and an inside-out outburst (right
        panel). The spectra are normalised to the local continuum. The
        lowermost graphs show a pure cold disc that evolves to full
        outburst (uppermost graph) over steps 1 to 5 of
        Tables~\ref{tab:outside-in} and \ref{tab:inside-out}, respectively.} 
      \label{fig:outburstsequence_norm}
    \end{figure*}

    The left panel of Fig.~\ref{fig:outburstsequence} shows the spectral
    evolution for this sequence from a pure cold disc to full outburst,
    as well as the left panel of Fig.~\ref{fig:outburstsequence_norm} where the
    spectra are normalised to the local continuum.
    The Balmer series turns from emission to absorption immediately after the
    outermost disk ring flipped into the hot state, because the overall
    disc flux in the optical is dominated by the flux of the outer rings
    and these are dominated by absorption in the hot state. 

    \begin{table}[h]
      \caption[]{Assembly of disc models for the simulated outside-in
    outburst.} 
      \label{tab:outside-in}
      \begin{center}
      \begin{tabular}{c c c c c c}
        \hline\hline\noalign{\smallskip}
        \# & Step 1 & Step 2 & Step 3 & Step 4 & Step 5\\
        \noalign{\smallskip}\hline\noalign{\smallskip}
         1 & 4.00cold & 4.00cold & 4.00cold & 4.00cold & 4.00cold\\
         2 & 6.00cold & 6.00cold & 6.00cold & 6.00cold & 6.00cold\\
         3 & 8.00cold & 8.00cold & 8.00cold & 8.00cold & 7.35hot\\
         4 & 9.00cold & 9.00cold & 9.00cold & 9.65hot  & 9.65hot\\
         5 & 10.0cold & 10.0cold & 13.5hot  & 13.5hot  & 13.5hot\\
         6 & 20.0cold & 21.0hot  & 21.0hot  & 21.0hot  & 21.0hot\\
         7 & 40.0hot  & 40.0hot  & 40.0hot  & 40.0hot  & 40.0hot\\
        \noalign{\smallskip}\hline
      \end{tabular}
      \end{center}
      \begin{quote}
      Numbers in ``step'' columns denote the radial position of the
      model in \(10^9\,\mathrm{cm}\), the following ``cold'' or ``hot'' label
      whether a cold or hot ring was used.
      \end{quote}
    \end{table}

    Similarly we modelled the inside-out outburst by a sequence of five
    disc models. In the first step the cold disc's innermost ring is
    replaced by the hot rings that lie inside of its radial position. At
    the same time the disc is extended inwards to the inner boundary of the
    hot disc at \(1\cdot10^9\mathrm{cm}\). For optical spectra, to which we
    will restrict our discussion in the following, this simplification can
    be justified because the inner disc rings will only contribute to the
    UV due to their high effective temperature and because they only cover
    a small surface area. In the subsequent steps, the next-neighbouring
    rings from the inside are replaced by hot ones. The complete assembly
    of the discs for the inside-out model sequence is shown in
    Table~\ref{tab:inside-out}.   

    \begin{table}[h]
      \caption[]{Assembly of disc models for the simulated inside-out
    outburst.} 
      \label{tab:inside-out}
      \centering
      \begin{tabular}{c c c c c c}
        \hline\hline\noalign{\smallskip}
        \# & Step 1 & Step 2 & Step 3 & Step 4 & Step 5\\
        \noalign{\smallskip}\hline\noalign{\smallskip}
         1  & 1.00hot & 1.00hot & 1.00hot & 1.00hot & 1.00hot\\
         2  & 1.10hot & 1.10hot & 1.10hot & 1.10hot & 1.10hot\\
         3  & 1.22hot & 1.22hot & 1.22hot & 1.22hot & 1.22hot\\
         4  & 1.35hot & 1.35hot & 1.35hot & 1.35hot & 1.35hot\\
         5  & 1.50hot & 1.50hot & 1.50hot & 1.50hot & 1.50hot\\
         6  & 1.66hot & 1.66hot & 1.66hot & 1.66hot & 1.66hot\\
         7  & 1.84hot & 1.84hot & 1.84hot & 1.84hot & 1.84hot\\
         8  & 2.05hot & 2.05hot & 2.05hot & 2.05hot & 2.05hot\\
         9  & 2.30hot & 2.30hot & 2.30hot & 2.30hot & 2.30hot\\
         10 & 2.60hot & 2.60hot & 2.60hot & 2.60hot & 2.60hot\\
         11 & 2.97hot & 2.97hot & 2.97hot & 2.97hot & 2.97hot\\
         12 & 3.43hot & 3.43hot & 3.43hot & 3.43hot & 3.43hot\\
         13 & 4.02hot & 4.02hot & 4.02hot & 4.02hot & 4.02hot\\
         14 & 4.80hot & 4.80hot & 4.80hot & 4.80hot & 4.80hot\\
         15 & 5.85hot & 5.85hot & 5.85hot & 5.85hot & 5.85hot\\
         16 & 6.00cold& 7.35hot & 7.35hot & 7.35hot & 7.35hot\\
         17 & 8.00cold& 8.00cold& 9.00cold& 9.65hot & 9.65hot\\
         18 & 9.00cold& 9.00cold& 10.0cold& 10.0cold& 13.5hot\\
         19 & 10.0cold& 10.0cold& 20.0cold& 20.0cold& 21.0hot\\
         20 & 20.0cold& 20.0cold& 40.0cold& 40.0cold& 40.0cold\\
         21 & 40.0cold& 40.0cold& -       & -       & -      \\
        \noalign{\smallskip}\hline
      \end{tabular}
    \end{table}

    The right panels of Figs.~\ref{fig:outburstsequence} and
    \ref{fig:outburstsequence_norm} show the spectral 
    evolution for this sequence from a pure cold disc to full outburst. In
    contrast to the outside-in scenario the hydrogen Balmer emission 
    only diminishes slowly during rise to outburst, while increasing
    absorption wings appear. 

    One has to keep in mind that the steps of our sequences are not
    equidistant in time. For an outside-in outburst, for example, the
    heating wave moves inwards quickly; and according to our models, the
    spectral lines change from emission to absorption as soon as a part of
    the outer region is in outburst, so one will observe an absorption-line
    spectrum during most of the rise of an outside-in outburst. In the case of an
    inside-out outburst, the heating wave moves rather slowly outwards. As
    our models show, the line spectrum of the disk does not change to pure absorption
    until the outermost regions are in outburst. Hence one would observe an
    emission-line spectrum most of the time of an inside-out outburst.

   \subsection{Decline}
    Decline from outburst to quiescence is mediated by a cooling wave. This
    wave always propagates outside-in as argued by
    e.g. Warner \cite{Warner1995}, so that the cooling can be studied by examining
    the inside-out moving heating wave of Table~\ref{tab:inside-out} in
    reverse order. According to the right panel of
    Fig.~\ref{fig:outburstsequence}, the hydrogen Balmer lines evolve
    smoothly from absorption to pure emission. 

  \section{Results and discussion}
  We used the models presented above to determine the nature of the
  outbursts in SS~Cyg by comparing them to spectra available in the
  literature. This turned out to be rather difficult, because adequate
  time-resolved spectra during rise to outburst are very rare even for the
  well-studied case of SS~Cyg. 

  Martinez-Pais et al. \cite{Martinez-Pais1996} presented time-resolved spectra for different
  outbursts of SS~Cyg. Among them are two spectra taken during a rise to an
  outburst of the symmetric type. According to their Fig.~2 the hydrogen
  Balmer emission lines decrease slowly between these two spectra, while
  the absorption wings increase like in the right panel of our
  Fig.~\ref{fig:outburstsequence} for the inside-out outburst. This leads
  to an identification of symmetric outbursts in SS~Cyg with inside-out
  outbursts, which is in good agreement to Smak's \cite{Smak1984a}
  description of type-B outbursts, but contrasts to the original
  conclusion of Martinez-Pais et al. \cite{Martinez-Pais1996}. They interpret the late
  appearance of the \ion{He}{ii} \(4686\,\AA\) line as a consequence of an
  outside-in propagating heating wave, because they assumed that the
  \ion{He}{ii} \(4686\,\AA\) line originates in the hot inner part of the
  disc. This is questionable in light of our models. The line should be
  significantly broader due to the higher Kepler velocity if this
  assumption is true. 

  ``Some indication of an increase in the hot spot's vertical size and,
  perhaps, brightness'' leads them to conclude further that  symmetric
  outbursts are connected with an instability of the secondary star, so
  they favoured the MTI for this outburst. This is put into question by
  our models, which indicate an inside-out outburst in favour of the DIM. 

  The observations of Clarke et al. \cite{Clarke1984} cover a complete
  outburst cycle of SS~Cyg. The outburst shows an asymmetric light
  curve. Three of its spectra before and during rise, as well as during
  maximum, are shown in Fig.~\ref{fig:SSCyg_outburst}. The spectral
  evolution of this outburst differs significantly from what was observed by
  Martinez-Pais et al. \cite{Martinez-Pais1996}: quiescent Balmer line
  emission abruptly disappears at the onset of rise to maximum, before full
  absorption sets in during the rise. This fits our outside-in model
  sequence, indicating that the asymmetric outbursts are indeed connected
  to outside-in, i.e.  type-A outbursts following Smak \cite{Smak1984a}. 

  The fact that the observed spectra of Clarke et al. \cite{Clarke1984} during full maximum
  do not show pure absorption lines like our model might be a consequence of the
  radial extension of the disc during outburst. This is not considered in our
  model, although it is expected in the DIM due to the higher transport of angular
  momentum during outburst. It might be possible that this portion of the disc
  outshines the basic inner part of the disc due to its large surface area. If
  this portion of the disc then has comparable properties to the current
  outermost grid point, which is relatively cool and emission-dominated, or at
  least continuum-dominated, the resulting model spectrum of the extended disc
  might show no absorption lines anymore.

  For decline from outburst to quiescence, another study by Hessman et
  al. \cite{Hessman1984} exists. Their Fig.~5 is in good agreement with the
  right panel 
  of our Fig.~\ref{fig:outburstsequence} if read from top to bottom, which
  means that they witnessed an outside-in propagating cooling wave. This 
  again agrees with the DIM. 

  During decline the \(4686\,\AA\) line of \ion{He}{ii} often shows a
  prominent emission feature like in the study of Hessman et
  al. \cite{Hessman1984}. This does not appear in our model. If it
  originates in the disc, it must arise from the inner parts, because only
  there does \(T_\mathrm{eff}\) become high enough to populate \ion{He}{ii}
  levels. Accordingly, the inner rings of our hot disc model show
  \ion{He}{ii} \(4686\,\AA\), however, not in emission but in
  absorption. This supports the  results of Unda-Sanzana et
  al. \cite{Unda-Sanzana2006}, which identified the gas stream/disc impact
  region as the origin of the \ion{He}{ii} \(4686\,\AA\) emission by means of
  Doppler tomography.  

  That Hessman et al. \cite{Hessman1984} observed central emission peaks in the
  Balmer lines right from the beginning of the decline might be indicating
  that not the complete disc but only the inner parts participated in the
  outburst. If for example the disc stays cold for 
  \(r>10\cdot10^{10}\,\mathrm{cm}\), we only have to compare the lower five
  curves of the right panel in Fig.~\ref{fig:outburstsequence} to the
  observation. As Hessman et al. \cite{Hessman1984} observed a short outburst without
  plateau, this would be again in good agreement with the DIM. There the
  short outbursts are attributed to discs that are not completely in
  outburst, while plateaus are supposed to appear if matter is accreted
  with a constant rate through a disc that is completely in outburst. 

  It will be interesting to extend our study to the UV range where especially
  the \ion{C}{iv} \(1550\,\AA\) line shows a similar behaviour to the Balmer
  lines.  For that purpose, heavier elements must be included in the model
  calculations, and the influence of the disc wind, which becomes obvious
  in the P~Cyg shaped profile of the \ion{C}{iv} \(1550\,\AA\) line in
  outburst, will be considered.

  Another question is the influence of metal opacities, which we
  have neglected here, on the hydrogen and helium lines. From our experience in
  working on stellar atmospheres (O and sdO stars), we would predict that metal
  line blanketing and surface cooling will produce slightly deeper H and
  He absorption lines in the hot disk; however, we expect no qualitative
  change in the optical spectrum. The situation is different in the cool
  ring models, which are optically thin. Work is in
  progress to investigate the metal line blanketing problem.

  \section{Summary}
  In this paper, we have presented NLTE model calculations for the accretion
  disc of SS~Cyg in outburst and quiescence. The resulting synthetic
  spectra describe the observed optical spectra and their transition well from
  absorption during outburst to emission during quiescence. 

  Simulations of the spectral evolution for outside-in and inside-out
  propagating heating waves were carried out. We compared them to published
  observations and conclude that symmetric outbursts belong to the
  inside-out type. This confirms DIM expectations (e.g. Smak 1984a),
  which are based on rise-time arguments and explicitly excludes the MTI
  model. In contrast, asymmetric outbursts seem to be outside-in
  outbursts.



\end{document}